\documentclass[reprint,aps,prl,superscriptaddress,showpacs]{revtex4-1}

\usepackage[letterpaper, margin=0.6in]{geometry}
\usepackage{amsmath,amssymb} 
\usepackage[]{graphicx}

\begin{document}
\title{Hanbury-Brown and Twiss anti-correlation in disordered photonic lattices}

\author{H. Esat Kondakci} 
\email[]{esat@creol.ucf.edu}
\author{Lane Martin}
\affiliation{CREOL, The College of Optics \& Photonics, University of Central Florida, Orlando, Florida 32816, USA}
\author{Robert Keil}
\affiliation{Institut f\"{u}r Experimentalphysik, Universit\"{a}t Innsbruck, Technikerstra$\beta$e 25, Innsbruck 6020, Austria}
\author{Armando Perez-Leija}
\affiliation{Max-Born-Institut, Max-Born-Stra$\beta$e 2A, 12489 Berlin, Germany}
\author{Alexander Szameit}
\affiliation{Institute of Applied Physics, Friedrich-Schiller-Universit\"{a}t Jena, 07743 Jena, Germany}
\author{Ayman F. Abouraddy}
\author{Demetrios N. Christodoulides} 
\author{Bahaa E. A. Saleh}
\affiliation{CREOL, The College of Optics \& Photonics, University of Central Florida, Orlando, Florida 32816, USA}

\begin{abstract}
We report measurements of Hanbury-Brown and Twiss correlation of coherent light transmitted through disordered one-dimensional photonic lattices. Although such a lattice exhibits transverse Anderson localization when a single input site is excited, uniform excitation precludes its observation. By examining the Hanbury-Brown Twiss correlation for a uniformly excited disordered lattice, we observe intensity anti-correlations associated with photon anti-bunching -- a signature of non-Gaussian statistics. Although the measured average intensity distribution is uniform, transverse Anderson localization nevertheless underlies the observed anti-correlation. 
\end{abstract}

\pacs{42.30.Ms, 42.25.Kb, 42.25.Dd} 
\maketitle
Incoherent light gains coherence upon free-space propagation, as dictated by the van Cittert-Zernike theorem \cite{Mandel1965a}.  The Hanbury-Brown and Twiss (HBT) interferometer \cite{HanburyBrown1956a,HanburyBrown1974} can reveal this acquired coherence by correlating intensity fluctuations at two different points. The HBT effect is a universal wave phenomenon that has been observed with free electrons \cite{Kiesel2002}, electrons in solid-state devices \cite{Oliver1999, Henny1999, Neder2007}, atoms in cold Fermi gases \cite{Rom2006, Jeltes2007a}, as well as interacting photons in nonlinear media \cite{Bromberg2010a}. In typical optical HBT scenarios, such as the original determination of the angular size of the star \textit{Sirius A} \cite{HanburyBrown1956a}, the radiation source is random while the medium transmitting the incoherent wave is deterministic. One might consider an alternative scenario in which a deterministic coherent input probes a scattering medium, which becomes itself the source of randomness. HBT measurements carried out on the emerging partially coherent light can provide insights into the nature of the disorder in the medium.

A particularly useful system for testing the impact of disorder on optical statistics is that of evanescently coupled waveguide arrays (or photonic lattices) with randomness introduced in the transverse direction  \cite{Christodoulides2003a}. This setting emulates time evolution of quantum-mechanical waves in time-independent disordered potentials. Indeed, by coupling a coherent input to a single lattice site, Anderson localization \cite{Anderson1958a} has been observed in the transverse direction upon ensemble averaging \cite{Raedt1989,Segev2013,Lahini2008a, Martin2011a}. Beyond the mean intensity observed in such experiments, unique features of the \textit{higher-order field correlations} involved in the HBT effect have only recently been explored \cite{Lahini2010a, Lahini2011a, Abouraddy2012b, Schlawin2012, DiGiuseppe2013, Gilead2015, Gneiting2016}. Indeed, HBT measurements can distinguish between the so-called `diagonal' and `off-diagonal' classes of lattice disorder, whereas such a delineation is not possible by observing the mean field alone \cite{Lahini2011a}. Furthermore, path-entangled photon pairs propagating along such lattices can emulate the quantum-mechanical waves associated with fermions and bosons \cite{Lahini2010a}, and can exhibit co-localization and anti-localization when the illumination is extended \cite{Abouraddy2012b, DiGiuseppe2013}.

In this letter, we report measurements of HBT interference in disordered photonic lattices excited \textit{uniformly} with an extended \textit{coherent} optical field. Light emerging from such a system is no longer coherent after ensemble averaging. The intensity fluctuations at \textit{any} site indicate a thermalization of optical statistics \cite{Kondakci2015, Kondakci2015b}. Here we measure the correlations between fluctuations at pairs of lattice sites. It is revealed -- surprisingly -- that at certain separations \textit{anti-correlations} emerge. This result implies that the optical field exiting the lattice is characterized by \textit{non}-Gaussian statistics that correspond to photon \textit{anti-bunching}. We argue that the mechanism underlying this behavior stems from the transverse localization of light, although localization itself is \textit{not} observed in the averaged intensity because the excitation is extended \cite{Kondakci2015}. Numerical simulations for both diagonal and off-diagonal disorder  when the input is uniform show no significant distinctions between the correlation functions in contrast to the single-site excitation case in \cite{Lahini2011a}. This is a clear indication that the excitation configuration plays an important role in shaping the correlation function. While off-diagonal disorder is associated with chiral symmetry, diagonal disorder is not. Nevertheless, the input excitation can help break chiral symmetry. Single site excitation maintains chiral symmetry while uniform excitation does not \cite{Kondakci2015b, Kondakci2016}.

The dynamics of optical propagation along disordered photonic lattices consisting of an array of evanescently coupled parallel waveguides is captured by a generic tight-binding model \cite{Christodoulides2003a}. The optical field is described by a set of coupled discrete Schr\"{o}dinger equations,
\begin{equation}
-i\frac{\mathrm{d}E_{x}}{\mathrm{d}z}=\beta_{x} E_{x}+C_{x,x-1} E_{x-1}+C_{x,x+1} E_{x+1};
\end{equation}
where $E_{x}$ is the complex field amplitude and $\beta_x$ is the propagation constant at the $x^{\mathrm{th}}$ site, and $C_{x,x+1}$ is the coupling coefficient between waveguides at lattice sites $x$ and $x+1$. \textit{Diagonal} disorder corresponds to constant coupling coefficients $C_{x,x+1}\!=\!\overline{C}$ and randomly selected propagation constants $\beta_x$, which is implemented by fixing the separations between waveguides of varying refractive index or width. On the other hand, \textit{off-diagonal} disorder corresponds to fixing the propagation constants $\beta_x\!=\!\overline{\beta}$ and randomly varying the coupling coefficients, which is realized by implementing random separations between identical waveguides. In the experimental results and numerical simulations reported here, we choose a uniform probability distribution for the random variables $C_{x,x+1}$ ($\beta_x$) of half-width $\Delta C$ ($\Delta\beta$) for off-diagonal (diagonal) disorder. All the disorder levels are scaled with the average coupling coefficient $\overline{C}$.

\begin{figure}[t]
\centering 
\includegraphics[width=8.6cm]{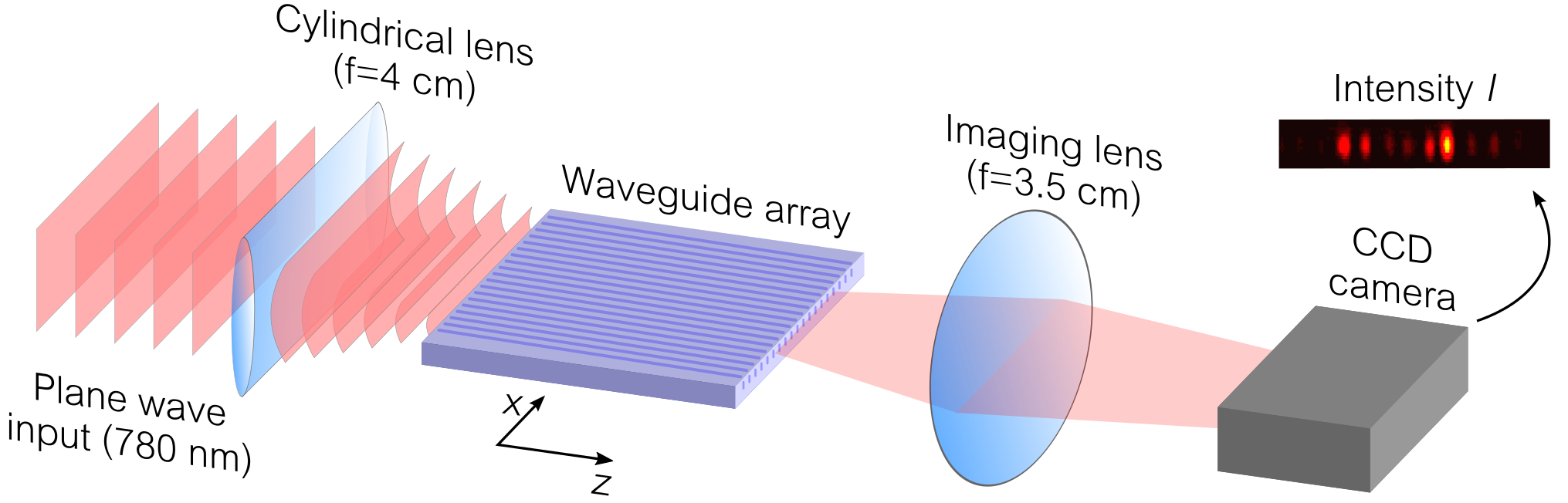} 
\caption{Experimental setup for HBT measurements. A cylindrical lens focusing along the vertical direction (focal length $f\!=\!4$~cm) couples an input coherent plane wave to the array. An imaging lens (focal length $f\!=\!3.5$~cm) magnifies the output facet of the waveguide array approximately fourfold. A CCD camera records the imaged output intensity distribution. An example of the recorded intensity $I$ is shown.} \label{Fig:setup}
\end{figure}

The normalized intensity correlation function at lattice sites $x$ and $x+\Delta x$ is 
\begin{equation}
g^{(2)}(x,x+\Delta x)=\frac{\langle I_x I_{x+\Delta x}\rangle}{\langle I_x\rangle\langle I_{x+\Delta x}\rangle},
\end{equation}
where $\langle\cdot\rangle$ denotes averaging over an ensemble of disorder realizations and $I_{x}\!=\!|E_{x}|^{2}$ (within a multiplicative constant). Assuming that the lattice disorder is statistically stationary in $x$ and that the input is uniformly extended, $g^{(2)}(\Delta x)$ depends only on the separation in coordinates $\Delta x$. Therefore, a `moving-average' approach can be implemented to produce an ensemble from a single disorder realization instead of repeating the measurement with multiple waveguide array samples (as confirmed in ~\cite{Martin2011a}).  This scheme is thus similar to the shifting-and-averaging utilized in the demonstration of transverse Anderson localization in diagonal \cite{Lahini2008a} and off-diagonal disordered lattices \cite{Martin2011a}, where the disorder was statistically stationary but the input was swept across single lattice sites.

In our experiment, we use a femtosecond-laser-written waveguide array consisting of 101 identical 4.9-cm-long waveguides \cite{Meany2015}; see Fig.~\ref{Fig:setup}. We implement off-diagonal disorder in the lattice by randomly varying the coupling coefficients between the neighboring waveguide pairs via a reliably calibrated control of their transverse separations -- as opposed to varying the refractive index of identical waveguides to implement diagonal disorder. The  average separation of the waveguides is 17~$\mu$m giving $\overline{C}\!\approx\!1.1$~cm$^{-1}$ at a wavelength of $\lambda\!=\!780$~nm. The coupling coefficients are drawn from a uniform probability distribution corresponding to a disorder level of $\Delta C\!=\!0.4$. A quasi-monochromatic plane wave at $\lambda\!=\!780$~nm from a laser diode is coupled to the waveguide array via a cylindrical lens focusing the beam along the vertical direction (focal length $f\!=\!4$~cm). Along the horizontal direction, the beam has a very extended Gaussian profile that is essentially exciting all waveguides in equal amplitudes. The output facet of the array is imaged to a CCD camera using a spherical lens ($f\!=\!3.5$~cm) with approximately $\times\!4$ magnification (Fig.~\ref{Fig:setup}).

\begin{figure}[t]
\centering  
\includegraphics[width=8.6cm]{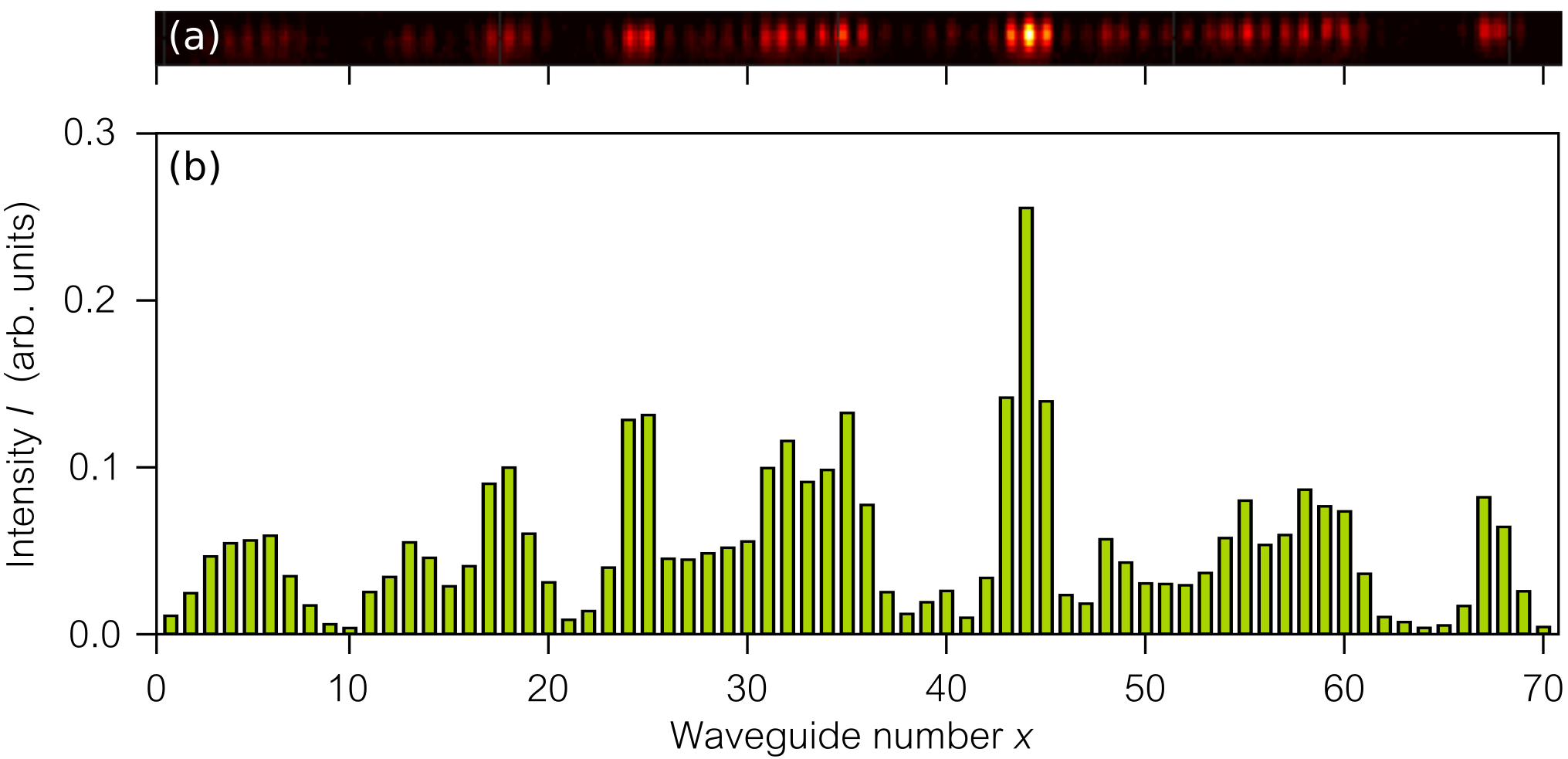} 
\caption{The output intensity distribution across the disordered waveguide array for  an extended uniform input excitation by a coherent optical field. (a) The CCD image of the output intensity distribution of the central 70 waveguides is depicted. Brighter colors imply higher intensities. (b) The bars represent the total intensity (in arbitrary units) for each lattice site obtained by \textit{vertically} integrating the intensity distribution depicted in (a), followed by binning the intensity in discrete cells along the \textit{horizontal} direction.}
\label{Fig:intensity}
\end{figure}

We plot in Fig.~\ref{Fig:intensity}(a) the measured intensity distribution across the output facet of the disordered waveguide array for an extended uniform excitation. To eliminate the effects of reflection from the lattice boundaries, only 71 waveguides in the lattice center are considered (removing 15 sites at both ends). These reflection effects from the boundaries are directly linked to ballistic expansion of the field in periodic lattices, which is linearly proportional to the product of the average coupling coefficient and the propagation distance, i.e., the normalized propagation distance $z\overline{C}$. In our case, $z\overline{C}\approx 5.4$ and this corresponds to expansion of the field across 15 lattice sites in both sides. Of course, the extended input excitation precludes an observation of localization in the random intensity distribution across the lattice. The bar plot in Fig.~\ref{Fig:intensity}(b) is obtained in two steps: integrating the CCD image along the vertical coordinate $y$ and then binning the intensity registered at each discrete site along $x$. A discrete speckle pattern $I_x$ clearly emerges across the lattice as predicted theoretically in ~\cite{Kondakci2015}.

We now move on to the analysis of the correlations in the experimental data. The two conditions for statistical stationarity discussed earlier are satisfied here: (1) the disorder is introduced into the waveguide array by \textit{independently} randomizing the coupling coefficients between each pair of neighboring waveguides; and (2) the excitation is uniformly extended. Consequently, the output field is statistically stationary and the correlation function depends only on the separation in coordinates $\Delta x$; that is, the intensity distribution is statistically invariant upon transverse translation. The large size of the lattice therefore suffices to produce an ensemble of disorder realizations from this single array. Exploiting this feature, we produce from the 71 lattice sites in Fig.~\ref{Fig:intensity}(b) an ensemble of size 40, each consisting of 31 lattice sites. 

\begin{figure}[b]
\centering 
\includegraphics[width=8.6cm]{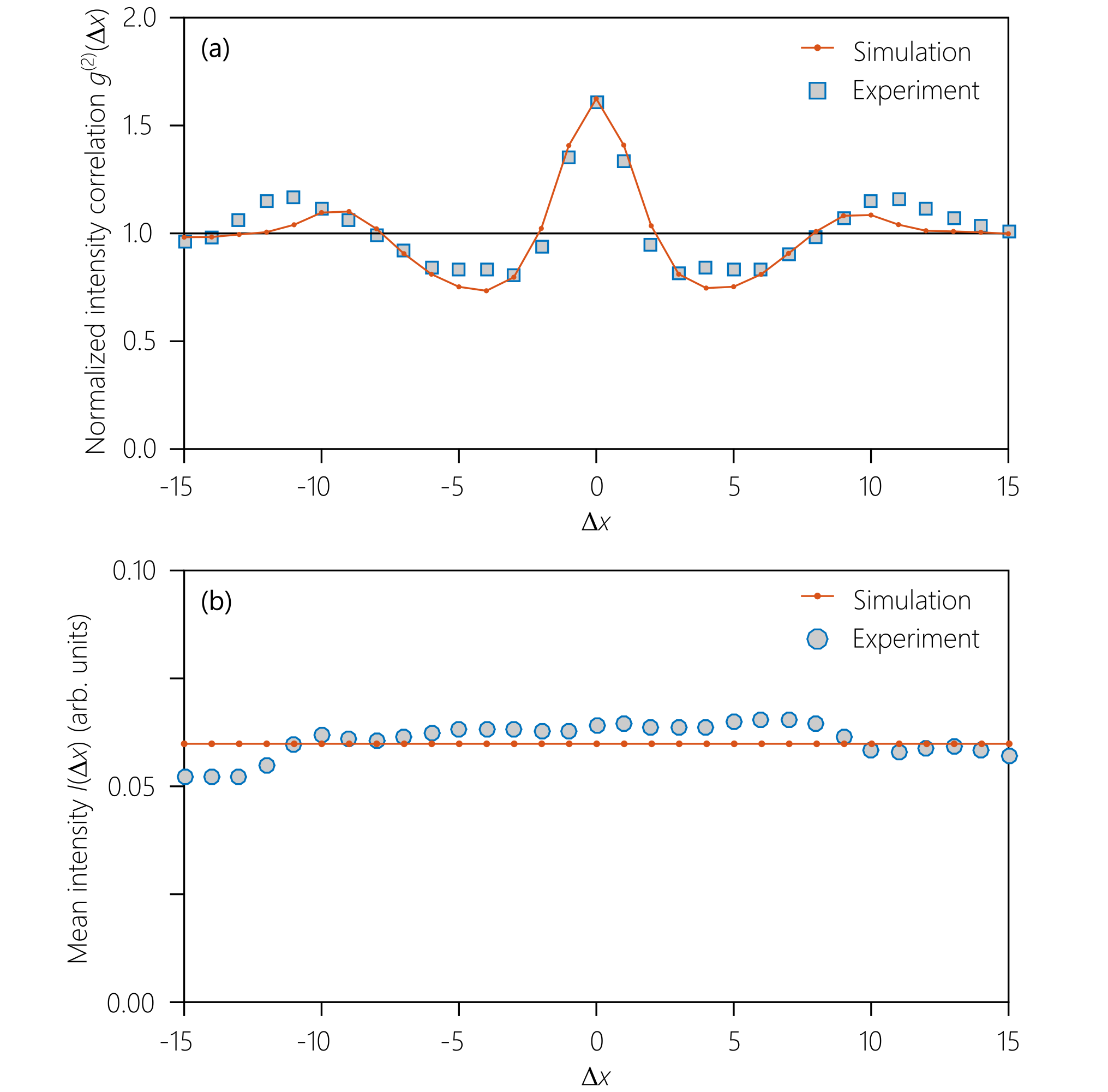}
\caption{Measured Hanbury-Brown and Twiss anti-correlation at the output of a disordered optical lattice. (a) The correlation function $g^{(2)}(\Delta x)$ as a function of the separation coordinate $\Delta x$. The signature for anti-correlation $g^{(2)}\!<\!1$ is observed in the region for which $\Delta x\!=\!2\!-\!7$. (b) The measured mean intensity in arbitrary units, corresponding to the same region in (a). Solid curves are theoretical predictions. The deviation in the tail of (a) the correlation function and (b) the mean intensity distribution from the theoretical predictions is attributed to the finite ensemble size (40 samples) utilized in the experiment.}
\label{Fig:HBT} 
\end{figure}

The correlation function $g^{(2)}(\Delta x)$, depicted in Fig.~\ref{Fig:HBT}(a), has a peak value of $g^{(2)}(0)\approx\!1.6$ at $\Delta x\!=\!0$. This result confirms that the emerging light is no longer coherent, but instead is partially thermalized \cite{Kondakci2015b}.  Surprisingly, $g^{(2)}(\Delta x)$ takes on values \textit{below unity} at waveguide separations in the range $\Delta x=2\!-\!7$. The stationarity of the output field is confirmed by computing the moving average of the intensity distribution with a window of size 30. The resulting average intensity $I(\Delta x)=\langle I_{x+\Delta x} \rangle_x=\frac{1}{40} \sum_{x=16}^{55} I_{x+\Delta x} $ is depicted in Fig.~\ref{Fig:HBT}(b) and exhibits uniform distribution confirming our prediction. Here $\langle\cdot\rangle_x$ denotes a moving-average over the lattice coordinate $x$. This is to be contrasted to the typical observation of Anderson localization due to single-site excitation. The numerical simulation shown in Fig.~\ref{Fig:HBT} with an ensemble size of $10^5$ for the disorder level and array length used in the experiment agrees well with the measurement. 

When considering \textit{thermal} light exhibiting Gaussian statistics and circularity \cite{Ohtsubo1976, Picinbono1994}, $g^{(2)}$ can be expressed in terms of the second-order field correlation $g^{(1)}$ through the Siegert relation \cite{Saleh1978},
\begin{equation}
g^{(2)}(\Delta x) = 1+|g^{(1)}(\Delta x)|^2,
\end{equation}
where $g^{(1)}(\Delta x)$ is the normalized field correlation 
\begin{equation}
g^{(1)}(x,x+\Delta x) = \langle E_x E^*_{x+\Delta x}\rangle / \sqrt{\langle I_x \rangle \langle  I_{x+\Delta x} \rangle}.
\end{equation}
Consequently, $g^{(2)}$ for thermal light takes values \textit{above} unity whenever the field displays correlation across the measurement points and is \textit{reduced} to unity in absence of any correlation. Measuring a value of $g^{(2)}$ below unity --- thus violation of the Siegert relation --- indicates anti-correlation (or negative covariance \cite{Saleh1978}) between the intensities at the measurement sites and, moreover, that the light is not truly thermal by virtue of its intensity fluctuations not obeying Gaussian statistics. Gaussian behavior is associated with satisfying the requirement of the central limit theorem, which requires the existence of a large number of statistically independent contributions. In the Anderson localization regime, this condition is no longer satisfied since only a small number of lattice eigenmodes are excited \cite{Kondakci2015b}. Consequently, the emerging light is characterized by non-Gaussian statistics.

\begin{figure}[b]
\centering 
\includegraphics[width=8.6cm]{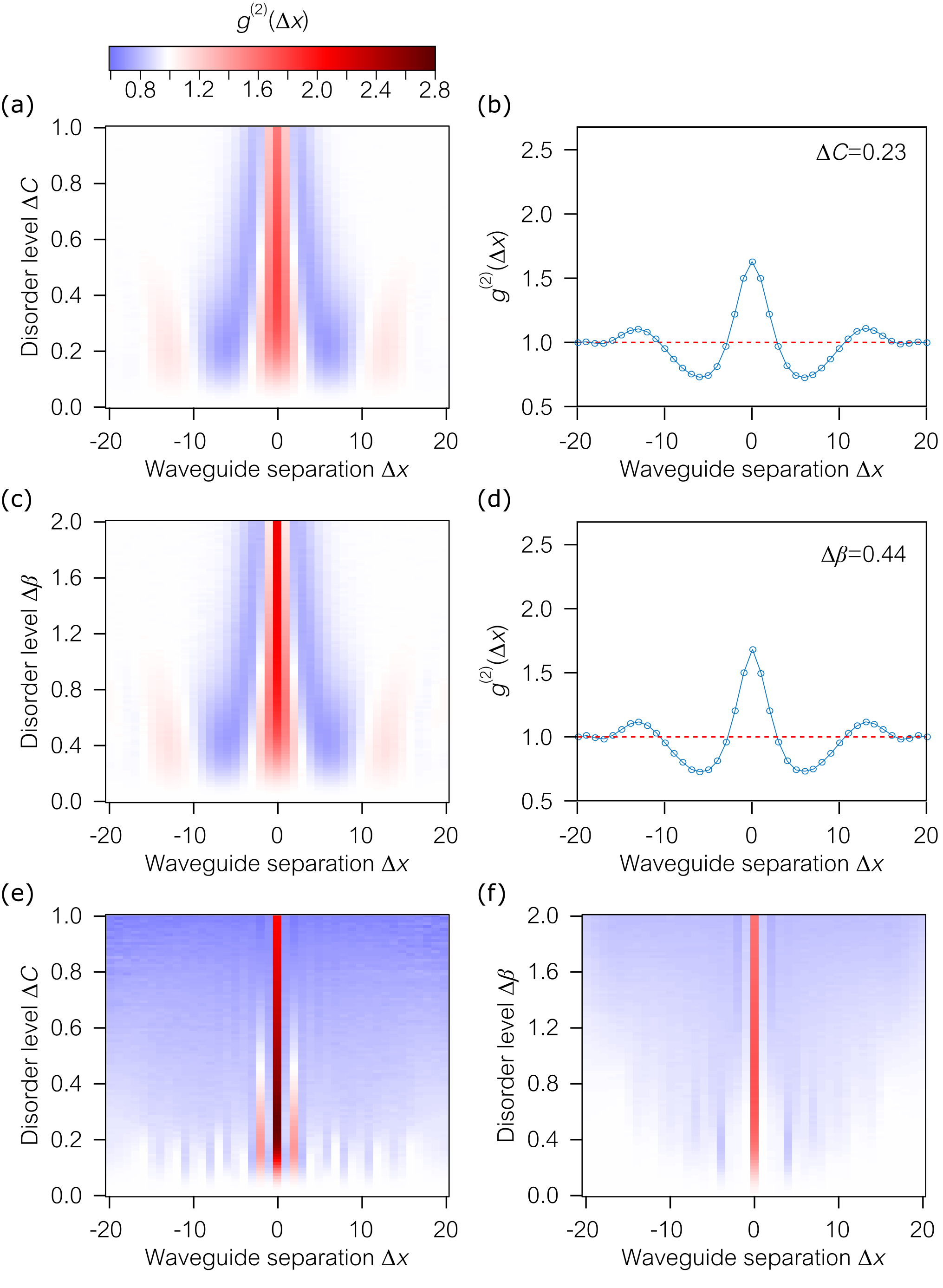}
\caption{Dependence of the Hanbury-Brown and Twiss correlation function $g^{(2)}(\Delta x)$ on the disorder class (off-diagonal or diagonal disorder), disorder level ($\Delta C$ or $\Delta\beta$), and excitation configuration. (a,b) In the 2D color plot we give $g^{(2)}(\Delta x)$ for off-diagonal disorder while varying the disorder level when the excitation is uniform ($E_x\!=\!1$). The line plot on the right depicts $g^{(2)}(\Delta x)$ that exhibits the maximum anti-correlation (minimum value of $g^{(2)}$) selected from the color plot on the left (the corresponding disorder level is given in the top-right corner). (c,d) Same as (a,b) for a lattice characterized by diagonal disorder.  (e)  $g^{(2)}(\Delta x)$ for a lattice characterized by off-diagonal disorder that is excited from a single point at the center ($E_x\!=\!\delta_{x,0}$). (f) Same as (e) for a lattice characterized by diagonal disorder. In all panels, the ensemble size is $10^5$ and $z\overline{C}\!=\!10$.}
\label{Fig:simulation} 
\end{figure}

A question may be raised whether the reported results in this experiment are dependent on the class of disorder used. To address this issue, we present numerical simulations of the correlation functions for arrays with off-diagonal and diagonal disorder as a function of disorder level. When the input field is uniform in amplitude and phase, there are no qualitative differences in $g^{(2)}$ between these disorder classes; compare Fig.~\ref{Fig:simulation}(a,b) to Fig.~\ref{Fig:simulation}(c,d). Both cases agree in the emergence of anti-correlations between sites having certain separations. On the other hand, upon single-site excitation the correlation functions show qualitatively different behavior as previously reported in ~\cite{Lahini2011a}. Here, $g^{(2)}$ exhibits an oscillatory pattern in off-diagonal disordered arrays (Fig.~\ref{Fig:simulation}(e)) which is absent when the disorder is diagonal (Fig.~\ref{Fig:simulation}(f)). 

These results can be understood by examining the underlying symmetries of the lattice in conjunction with the excitation configuration. It is now understood that lattices with off-diagonal disorder feature a disorder-immune `chiral symmetry' which results in the lattice eigenmodes and eigenvalues occurring in skew-symmetric pairs. This symmetry is absent in lattices with diagonal disorder. The unique consequences of chiral symmetry -- such as the emergence of super-thermal light from a coherent input -- become dormant when the mode pairs are excited asymmetrically. A single-site excitation maintains chiral symmetry, while uniform lattice excitation breaks this symmetry, a condition that renders the diagonal and off-diagonal disorder classes essentially similar to each other. This explains the sub-thermal statistics observed here $g^{(2)}(0)\!<\!2$ despite the off-diagonal disorder, which is broken here by virtue of the uniform excitation. It is an open question how the input excitation and lattice disorder can be designed together to maximize the intensity anti-correlation that can be produced by a disordered photonic lattice.

In conclusion, we have demonstrated experimentally that the propagation of a uniformly extended coherent field in disordered lattices results in the emergence of intensity anti-correlations -- photon anti-bunching -- at certain waveguide separations. This anti-correlation is verified by observing the normalized intensity correlations drop below unity at these separations and is supported by simulations of the propagation dynamics. This type of violation of Siegert's relation implies a departure of the field statistics at the output from Gaussianity (departure from circularity is accompanied enhanced correlation). Underlying this anti-correlation is the transverse localization of light, although the localization itself is not observed because the excitation is extended. Since a uniform excitation configuration breaks the chiral symmetry by exciting the chiral modes with unequal amplitudes, off-diagonal and diagonal disorder behave in similar manners. 

\bibliography{library_short,books}
\end{document}